# *Data article*

**Title: Data on the annual aggregated income taxes of the Italian municipalities over the quinquennium 2007-2011.**

**Authors: Marcel Ausloos, Roy Cerqueti, Tariq A. Mir**

**Affiliations:**
<u>Marcel Ausloos:</u> Institute of Accounting and Finance, School of Management,
University of Leicester, University Road,
Leicester, LE1 7RH, UK
<u>Roy Cerqueti:</u> University of Macerata, Department of Economics and Law,
via Crescimbeni 20, I-62100, Macerata, Italy
<u>Tariq A. Mir:</u> Nuclear Research Laboratory, Astrophysical Sciences Division,
Bhabha Atomic Research Center, Srinagar-190 006,
Jammu and Kashmir, India

**Contact email: roy.cerqueti@unimc.it**

**Abstract**
*This dataset contains the annual aggregated income taxes of all the Italian municipalities over the years 2007-2011. Data are clustered over the Italian regions and provinces. The source of the data is the Italian Ministry of Economics and Finance. The administrative variations in Italy over the quinquennium have been taken into account. Data are useful to understand the economic structure of Italy at the microscopic level of municipalities. They can serve also for making comparisons between economical aspects and other features of the Italian cities.*

**Specifications Table**

| | |
|---|---|
| Subject area | *Economics* |
| More specific subject area | *Tax income, GDP* |
| Type of data | *Table* |
| How data was acquired | *Research center of the Italian Ministry of Economics and Finance* |
| Data format | *Raw* |
| Experimental factors | *The administrative changes have been taken into account, and data have been pretreated in this sense (see the details below)* |
| Experimental features | *The merging of cities have been treated by aggregating the data of the merged municipalities* |
| Data source location | *Italy* |
| Data accessibility | *Data are included in this article as a supplementary material in a Microsoft Excel Worksheet* |
| Related research article | *M. Ausloos, R. Cerqueti, T.A. Mir (2017) Data science for assessing possible tax income manipulation: the case of Italy, Chaos, Solitons and Fractals 104:238-256* |

**Value of the data**

- *Data source is an official Institutions, the Italian Ministry of Economics and Finance*
- *Data are complete for all the quinquennium 2007-2011 and for all the Italian municipalities*
- *Data have been treated to include all the administrative variations in Italian cities and provinces occurred over the quinquennium*
- *The dataset has been widely employed in scientific papers written by the authors (see references [1-9])*

**Data**

*This article is associated to a Microsoft Excel Worksheet as a supplementary material. The file has 20 data sheets corresponding to the 20 Italian regions of Italy. In each sheet, the regional data for the years 2007-2011 is presented. Sheets are named after the regions of which they contain the data. Each municipality has a specific reference to the Italian province (in brackets, the acronyms of the provinces). The background of the study relies to the economic analysis of Italy at the microlevel of cities contributions to the national GDP for a relevant time period, along with the related implications on the understanding on the socio-economic aspects of Italian reality.*

**Experimental Design, Materials and Methods**

*Data have been arranged for including the changes in the Italian administrative structure occurred over the quinquennium 2007-2011.*
*Indeed, Italy is composed of 20 regions, more than 100 provinces and more than 8000. We enter soon the precise values.*
*Municipalities are included in provinces, and provinces belong to regions. Such territories do not intersect.*
*The number of regions has been equals to 20 over the quinquennium. Differently, the number of cities has been modified as follows : 8101, 8094, 8094, 8092, 8092, - from 2007 till 2011, respectively.*
*The dataset claims a number of 103 provinces in 2007, with the addition of seven new provinces after 2007 (BT, CI, FM, MB, OG, OT, VS). In this respect, we note a discrepancy*
*between data and the analysis of the administrative provinces evolution. Specifically, the four provinces CI, MB, OG, OT have been created by a regional law in July 12th, 2001 and only lin 2005 they became operative. Differently, BT, FM and VS have been instituted on June 11th, 2004 and became operative only on June 2009.*
*This discrepancy has been solved by taking as official the original data provided by the Italian Ministry of Economics and Finance, so that the number of provinces is 103, 110, 110,*
*110, 110 - from 2007 till 2011, respectively.*
*A correct use of the dataset imposes the comparison of identical lists. We take as reference year 2011, with 110 provinces and 8092 municipalities. To uniform the data, we have considered the merging of cities of the inclusion of some of them into previously existing municipalities. Here it is the list:*
*(i) CAMPOLONGO AL TORRE (UD) and TAPOGLIANO (UD) have*

*merged after a public consultation, held on Novembre 27th, 2007, into*
*CAMPOLONGO TAPOGLIANO (UD).*
*(ii) LEDRO (TN) is a new municipality created by merging (after a public consultation,*
*held on Novembre 30th, 2008) BEZZECCA (TN), CONCEI (TN),*
*MOLINA DI LEDRO (TN), PIEVE DI LEDRO (TN), TIARNO DI*
*SOPRA (TN) and TIARNO DI SOTTO (TN).*
*(iii) COMANO TERME (TN) is a new municipality coming out from the merging of BLEGGIO INFERIORE*
*(TN) and LOMASO (TN), by a regional law of*
*November 13th, 2009.*
*(iv) CONSIGLIO DI RUMO (CO) and GERMASINO (CO) were annexed*
*by GRAVEDONA (CO) on May 16th, 2011 and February 10th, 2011. The resulting new municipality is*
*GRAVEDONA ED UNITI (CO).*
*The implications of the study are relevant and very intuitive. According to the considered dataset, researchers are able to have a panoramic view at the microscopic level of cities of the Italian economic situation over a remarkable time horizon. Since administratively regions and provinces are collections of municipalities and such an information is explicitly given in the dataset, then the data are useful also for having a complete analysis of the economic structure of Italy at provincial and regional levels.*

**References**

[1] M. Ausloos, R. Cerqueti (2015) Socio-economical analysis of Italy: The case of hagiotoponym cities, The Social Science Journal, 52(4):561-564.

[2] M. Ausloos, R. Cerqueti (2016) Religion-based Urbanization Process in Italy: Statistical Evidence from Demographic and Economic Data, Quality and Quantity, 50(4):1539-1565.

[3] M. Ausloos, R. Cerqueti (2017) Intriguing yet simple skewness - kurtosis relation in economic and demographic data distributions, pointing to preferential attachment processes, Journal of Applied Statistics, doi:10.1080/02664763.2017.1413077.

[4] M. Ausloos, R. Cerqueti, T.A. Mir (2017) Data science for assessing possible tax income manipulation: The case of Italy. Chaos, Solitons & Fractals, 104:238-256.

[5] R. Cerqueti, M. Ausloos (2015a) Cross Ranking of Cities and Regions: Population vs. Income, Journal of Statistical Mechanics: Theory and Experiments, 7:P07002.

[6] R. Cerqueti, M. Ausloos (2015b) Evidence of Economic Regularities and Disparities of Italian Regions From Aggregated Tax Income Size Data. Physica A: Statistical Mechanics and its Applications,421(1):187-207.

[7] R. Cerqueti, M. Ausloos (2015c) Statistical Assessment of Regional Wealth Inequalities: the Italian Case. Quality and Quantity, 49(6):2307–2323.

*[8] R. Cerqueti, M. Ausloos (2016) Studies on Regional Wealth Inequalities: the Case of Italy, Acta Physica Polonica A 129(5):959-964.*

*[9] T.A. Mir, M. Ausloos, R. Cerqueti (2014) Benford's law predicted digit distribution of aggregated income taxes: the surprising conformity of Italian cities and regions. The European Physical Journal B, 87:261.*